\documentclass[twocolumn]{aastex63}
\usepackage{amsmath}

\shorttitle{FRBs from Magnetars Formed from NSWD Mergers}
\shortauthors{Zhong \& Dai}

\begin{document}
\title{Magnetars from Neutron Star--White Dwarf Mergers: Application to Fast Radio Bursts}
\author[0000-0002-1766-6947]{Shu-Qing Zhong}
\affil{School of Astronomy and Space Science, Nanjing University, Nanjing 210093, China; sqzhong@hotmail.com, dzg@nju.edu.cn}
\affil{Key laboratory of Modern Astronomy and Astrophysics (Nanjing University), Ministry of Education, Nanjing 210093, China}
\author[0000-0002-7835-8585]{Zi-Gao Dai}
\affil{School of Astronomy and Space Science, Nanjing University, Nanjing 210093, China; sqzhong@hotmail.com, dzg@nju.edu.cn}
\affil{Key laboratory of Modern Astronomy and Astrophysics (Nanjing University), Ministry of Education, Nanjing 210093, China}

\begin{abstract}
It is widely believed that magnetars could be born in core-collapse supernovae (SNe), binary neutron star (BNS) or binary white dwarf (BWD) mergers,
or accretion-induced collapse (AIC) of white dwarfs. In this paper, we investigate whether magnetars could also be produced from neutron star--white dwarf (NSWD) mergers, motivated by FRB 180924-like fast radio bursts (FRBs) possibly from magnetars born in BNS/BWD/AIC channels suggested by \cite{mar19}. By a preliminary calculation, we find that NSWD mergers with unstable mass transfer could result in the NS acquiring an ultra-strong magnetic field via the dynamo mechanism due to differential rotation and convection or possibly via the magnetic flux conservation scenario of a fossil field. If NSWD mergers can indeed create magnetars, then such objects could produce at least a subset of FRB 180924-like FRBs within the framework of flaring magnetars, since the ejecta, local environments, and host galaxies of the final remnants from NSWD mergers resemble those of BNS/BWD/AIC channels. This NSWD channel is also able to well explain both the observational properties of FRB 180924-like
and FRB 180916.J0158+65-like FRBs within a large range in local environments and host galaxies.
\end{abstract}

\keywords{Compact binary stars (283); Gravitational waves (678); Magnetars (992); Radio bursts (1339)}

\section{Introduction}
\label{sec:introduction}
Fast radio bursts (FRBs) have remained an extragalactic enigma so far \citep{katz18,popo18,petr19,cor19}
since they were discovered by \cite{lorimer07}, \cite{kean12}, and \cite{tho13}.
They are millisecond-duration coherent radio pulses with average upper limits of the peak luminosity
$L_{\rm p}\sim1\times10^{42}-8\times10^{44}~{\rm erg~s^{-1}}$ and
energy $E\sim7\times10^{39}-2\times10^{42}~{\rm erg}$ \citep{zhang18},
characterized by a single peak mainly or multiple peaks rarely \citep{cham16,far18,pro19},
phenomenally divided into repeating bursts \citep{spi16,chime19a,chime19b,kumar19,chime20} and non-repeating bursts.
To date, over 100 FRBs have been reported in the literature and collected in the FRB Catalogue\footnote{http://www.frbcat.org} \citep{petr16}.
Meanwhile, to explain this radio phenomena, dozens of progenitor models involved in
compact objects have been proposed \citep{kash13,fal14,lyub14,geng15,dai16,gu16,wang16,zhang14,mur16,zhang16,zhang17,lyut16,met17,belo17,deng18,mar18b,met19,belo19}, accounting for  non-repeating and/or repeating FRBs.
A full model list can refer to \cite{plat19}\footnote{http://frbtheorycat.org}.

One interesting group of models relevant to a young flaring magnetar with single or clustered flares have been proposed to
give rise to non-repeating or repeating bursts, respectively \citep{lyub14,katz16,belo17,belo19,met17,kumar17,lu18,met19}.
One of them has been developed to successfully explain nearly all observational properties of FRBs
such as polarization, rotation measure \citep[RM;][]{mich18}, frequency downward drift \citep{hes19},
persistent radio source and optical counterpart \citep{chat17,ten17}, circum-burst dispersion measure (DM),
and the ``dark periods'' between bursts and clustered burst arrival times appearing in FRB 121102
and its hosted low metallicity dwarf star forming galaxy and its surrounding dense,
highly magnetized, and dynamic plasma environment \citep{chat17,ten17}
within the framework of synchrotron maser emission from decelerating relativistic blast waves
produced by flare ejecta from young magnetars \citep{met19}. In this framework,
repeating FRBs similar to FRB 121102 are arise from young and very active millisecond magnetars
quite possibly connected with superluminous supernovae (SLSNe)
and long-duration gamma-ray bursts \citep[LGRBs;][]{met17}.
Therefore, young magnetars giving rise to FRB 121102-like FRBs might be formed during the core-collapse of massive stars
associated with SLSNe or LGRBs (SLSNe/LGRBs channels).

On the other hand, young millisecond magnetars could also be born
in binary neutron star (BNS) mergers \citep{ross03,price06,gia13},
binary white dwarf (BWD) mergers \citep{king01,yoon07,sch16},
or accretion-induced collapse (AIC) of
white dwarfs \citep[WD;][]{nomo91,tau13,sch15}. These magnetars
could produce FRBs analogous to FRB 180924 \citep{ban19}, as suggested by \cite{mar19}.
Compared with FRBs created from magnetars born in SLSNe/LGRBs channels,
the FRBs produced from magnetars born in BNS/BWD/AIC channels could
have a distinct observational properties.
Just like FRBs 180924 and 190523 \citep{ravi19a}, likely as well as FRB 181112 \citep{pro19},
they host an old massive galaxy with a relatively low rate of star formation and relatively high metallicity,
lie in a large spatially offset location relative to the central containment region of the galaxy, and
have low DM and RM contributions from the host galaxy
and no bright persistent radio source \citep{ban19,ravi19a}.
If this is the case, FRBs could be divided into two populations:
FRB 121102-like bursts stem from young magnetars born in SLSNe/LGRBs channels,
while FRB 180924-like cases come from those young magnetars
born in BNS/BWD/AIC channels. Additionally, most FRB 180924-like bursts should also be repeating
in the flaring magnetar framework due to the event rate comparison between magnetars
and total FRB events \citep{nich17,mar19,ravi19b},
which is supported by FRB 171019 followed by faint bursts \citep{kumar19}.

In this paper, we investigate whether or not FRB 180924-like bursts are also likely to be generated
by magnetars born in an alternative possible channel: neutron star--white dwarf (NSWD) mergers.
This channel has also been briefly mentioned and/or discussed
by \cite{liu18,liu20}, \cite{kho19}, and \cite{belo19} previously.
To answer this question, we need to study (1) whether this channel can form magnetars or not,
and (2) if it can, whether the formed magnetars can account for the observations of FRB 180924-like bursts
in the flaring magnetar framework. If this channel can indeed form magnetars,
it could increase the magnetar event rate to some extent and contribute to at least a subset of FRBs
similar to FRBs 180924, 190523, 181112, and even 180916.J0158+65. In the following,
we organize the structure of the paper: \S \ref{sec:NS-WD} introduces the possibility
and speculation that NSWD mergers could form magnetars;
whether or not the NSWD channel can explain the observations of FRB 180924-like cases
is discussed in \S \ref{sec:explanation};
and a summary and discussion are presented in \S \ref{sec:summary}.

\section{Magnetars from NSWD Mergers}
\label{sec:NS-WD}
The explosive outcomes of NSWD mergers have been explored
in the literature \citep{met12,mar16,mar17,zen19a,zen19b,fern19},
but the final remnants of these events have been little investigated
\citep[see][]{pas11a,pas11b,mar16}.
Generally, there are two evolutionary pathways for NSWD binaries,
which depend on the critical mass ratio $q_{\rm crit}=M_{\rm WD,crit}/M_{\rm NS}$,
where $M_{\rm WD,crit}$ is the critical WD mass and $M_{\rm NS}$ is the NS mass.
The first pathway is that the WD fills its Roche lobe and its matter undergoes
stable mass transfer to the NS if $q<q_{\rm crit}$, evolving into
an ultra-compact X-ray binary. The second pathway is that the WD is tidally disrupted
by the NS via unstable mass transfer on a rather short dynamical timescale for $q>q_{\rm crit}$,
leading to an NSWD merger \citep{hje87,hur02}.
The critical mass ratio is related to the critical WD mass $M_{\rm WD,crit}$, which is found
to be $M_{\rm WD,crit}=0.37~M_{\odot}$ \citep{van12}
or $M_{\rm WD,crit}=0.2~M_{\odot}$ \citep{bob17}.
Thus an NSWD merger with $q>q_{\rm crit}$ is in the case of unstable mass transfer.
Toonen et al. (2018) pointed out that over 99.9\% of semi-detached NSWD systems
would merge when $M_{\rm WD,crit}=0.2~M_{\odot}$,
which indicates that the NSWD merger is a prevalent fate of semi-detached NSWD binaries.
After an NSWD merger, as shown by \cite{pas11a,pas11b}, the final remnant both
in the inspiraling case and in the head-on case is a spinning Thorne-Zytkow-like object
\citep[TZIO;][]{thor77} surrounded by a massive extended hot disk composed of WD debris
without explosive outcomes. Considering the disk winds and nuclear burning,
\cite{mar16} suggested an NSWD merger likely forms an isolated millisecond pulsar surrounded
by an accretion disk at the final stage. Whether or not these final remnants evolve into magnetars
has received less attention. How the magnetic fields of the final remnants evolve
remains unknown. Fortunately, it has been suggested that the ultra-strong magnetic fields
in magnetars may result from two main scenarios \citep[for a review see][]{tur15}.
Although these two scenarios are mainly used for nascent NSs born in SLSNe/LGRBs/BNS/BWD/AIC channels,
we guess that they might also be used for ``renascent''
(magnetic field undergoes amplification) NSs formed in the NSWD channel.

\subsection{$\alpha$--$\omega$ Dynamo}
\label{subsec:dynamo}
The first scenario that we consider is magnetic field amplification by a vigorous dynamo action at the early,
highly convective stage of the NSs after mergers: the $\alpha$ dynamo arising
from the coupling of convective motions and rotation, and the $\omega$ dynamo
driven by differential rotation \citep{dun92,thom93}.
For an NSWD merger, the magnetic field of the NS remnant surrounded by
a massive disk may either increase via a dynamo winding-up process \citep[as suggested by][]{pas11b}
or decrease through an enhanced ohmic dissipation of accreted matter
in the NS' crust \citep{urp97,konar97,cum04},
somewhat similar to the finding of \cite{sun19}.
We discuss whether this finding is true in the following.
\cite{bob17} showed that NSWD mergers exhibit an exponentially increasing rate of
mass transfer during different phase transitions \citep[see Figure 12 in][]{bob17},
and the NSWD mergers in which the WDs have a higher mass would have a shorter
dynamical timescale between the onset of significant mass transfer and the final
merger \citep[e.g., only $t_{\rm dyn}\sim10^{-3}(10^{-5})~{\rm yr}\sim3\times10^4(3\times10^2)~{\rm s}$ for WD mass $\sim0.75(1.2)~M_{\odot}$, see Figure 13 in][]{bob17}.
However, it is not true that the maximal rate of disk accretion onto the final NS in
Figures 11 and 12 of \cite{bob17} is limited by the Eddington rate
since the disk accretion of NS can be highly super-Eddington.
If the majority of mass is lost via a disk wind or possible a jet in the mass transfer process
and only $0.05~M_{\odot}$ can be accreted onto the NS, as discussed in \cite{mar16}
for a WD with mass $0.6~M_{\odot}$ (close to $0.75~M_{\odot}$),
the NS would accrete the disk material onto its surface with
an average rate $\dot{M}\sim10^{-6}~M_{\odot}~{\rm s}^{-1}$
during the short dynamical timescale $\sim3\times10^4$ s. In this case,
the accretion of the final NS surrounded by a massive disk may
let it differentially and rapidly rotate, as possibly shown by the simulation results of \cite{pas11b},
even if \cite{pas11b} did not take into account explosive outcomes.
Moreover, during this short dynamical timescale,
transient ohmic dissipation of the final NS could be ignored,
see an estimate in \cite{sun19} and Equation (9) of \cite{urp97}.
Therefore, we just need to focus on the magnetic field amplification
of the final NS that accretes the WD debris material from the disk during the merger.
Owing to the lack of previous studies of the magnetic field evolution of NSWD mergers
with unstable and rapid mass transfer, we perform only a preliminary analysis
on the $\alpha-\omega$ dynamo induced by possible differential rotation
and/or convection that can amplify the NS' magnetic field during the NSWD mergers.
In the final paragraph of this subsection, we would also discuss the dynamo induced by
the magneto-rotational instability \citep[MRI;][]{bal98} in the disk that also
might contribute to the NS' magnetic field amplification.

\begin{figure}
\includegraphics[width=0.5\textwidth, angle=0]{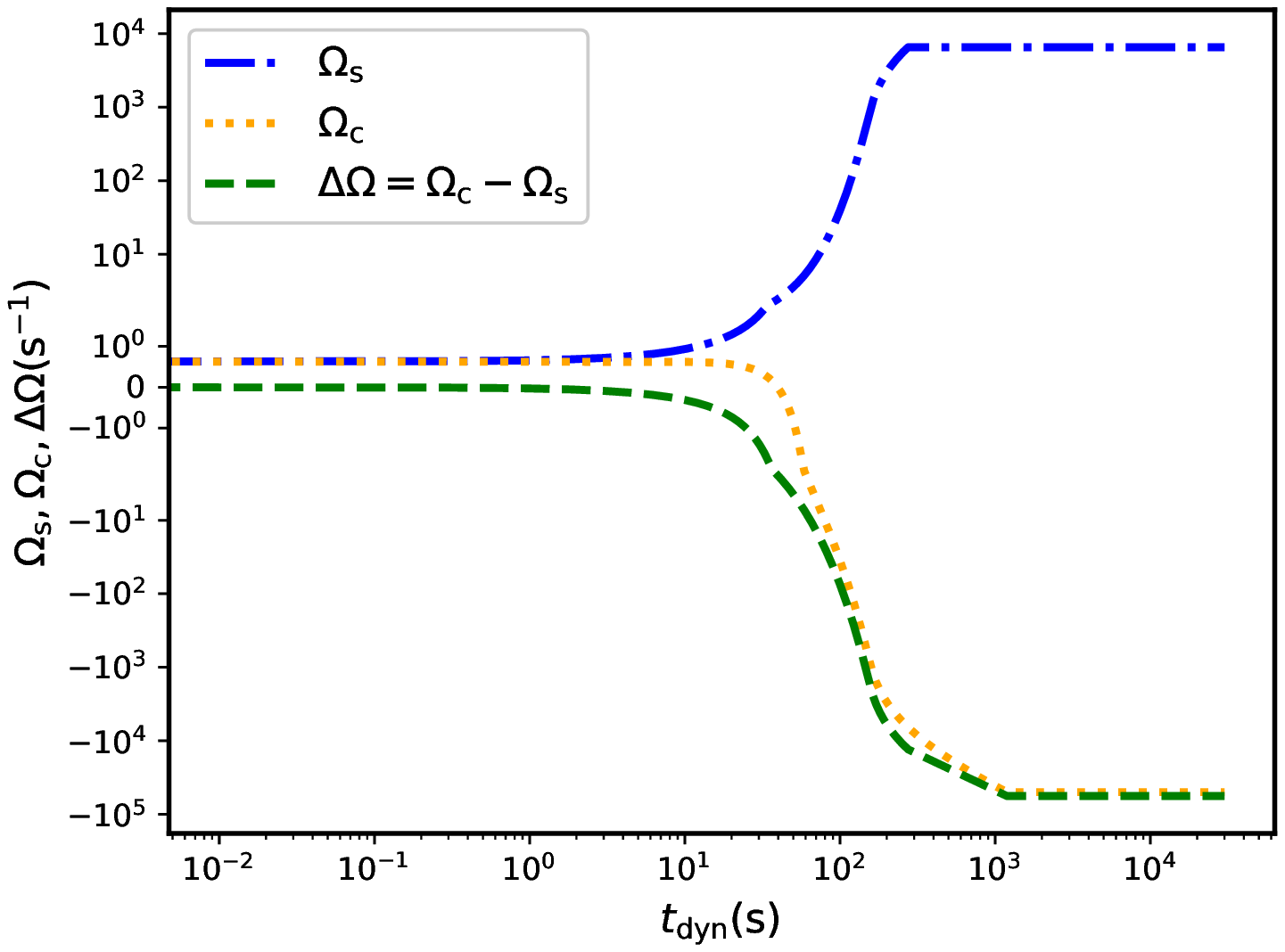}
\includegraphics[width=0.5\textwidth, angle=0]{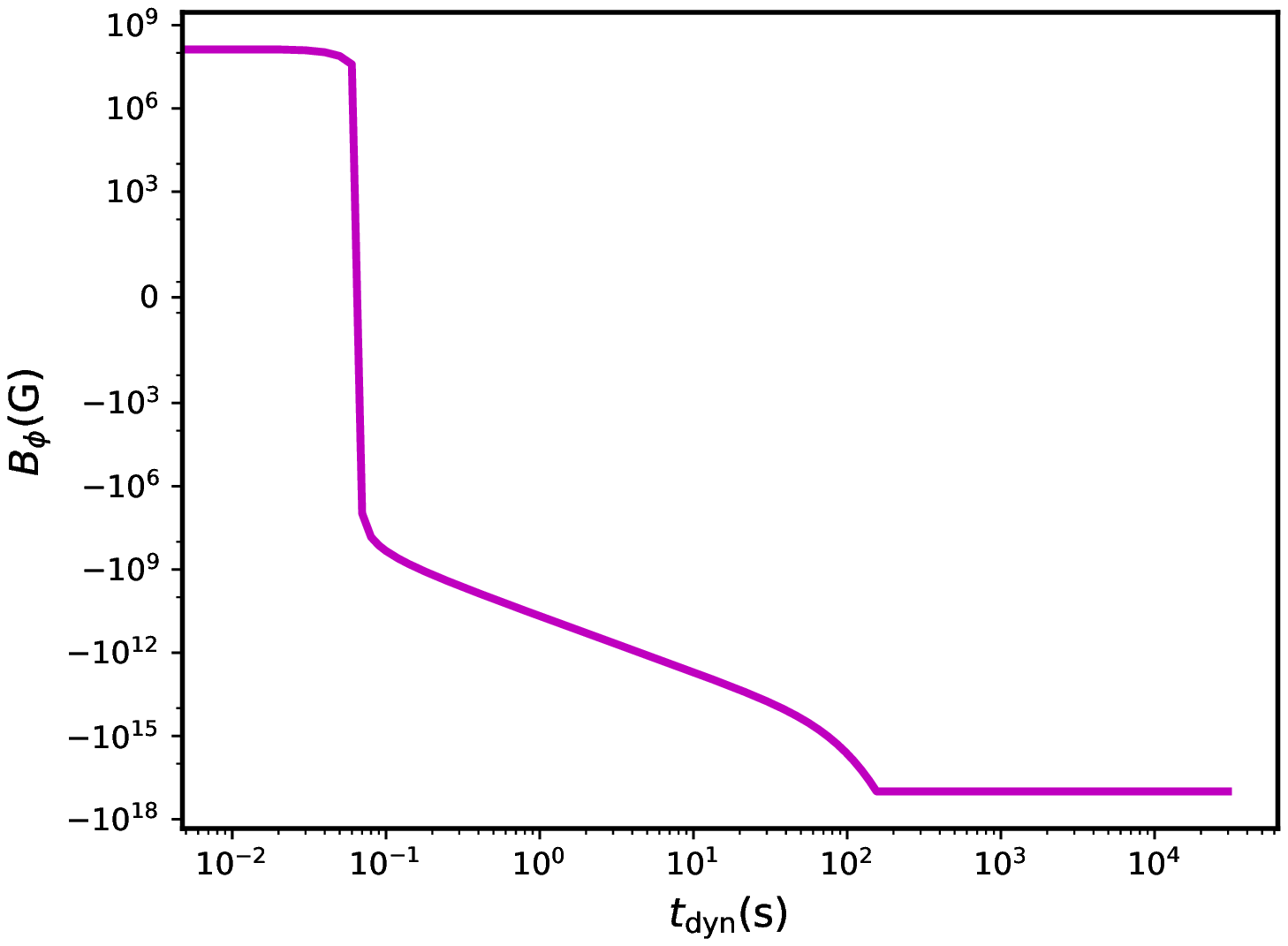}
\caption{Evolution of the angular velocities of the shell and the core and the differential angular velocity $\Delta\Omega$ (top panel), and the torodial magnetic field $B_{\phi}$ (bottom panel) of the NS
in an NSWD merger during the dynamical timescale $t_{\rm dyn}=3\times10^4~{\rm s}$.}
\label{fig:solution}
\end{figure}

\begin{figure}
\includegraphics[width=0.5\textwidth, angle=0]{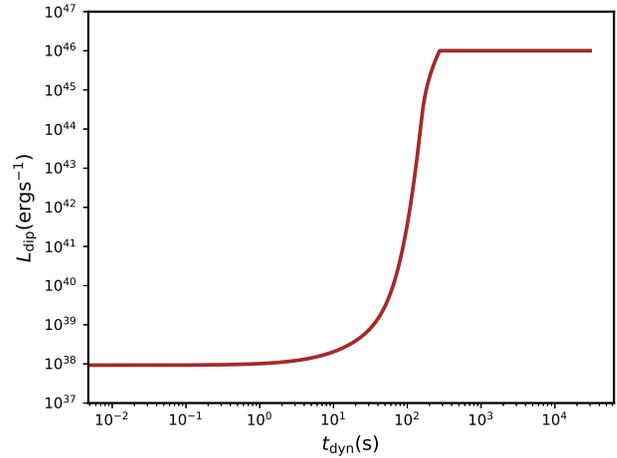}
\caption{Magnetic dipole radiation evolution of the NS in an NSWD merger
during the dynamical timescale.}
\label{fig:Ldip}
\end{figure}

We assume that the final NS with differential rotation induced by accretion has two components:
the core and the shell divided by a boundary at the radius $R_{\rm c}\cong0.5R$
(where $R$ is the NS radius), as done by \cite{dai06},
its accretion can be generally determined by the Alfv\'{e}n radius
\begin{eqnarray}
r_{\rm m}&=&(B_{\rm s}R^3)^{4/7}(GM)^{-1/7}\dot{M}^{-2/7} \nonumber\\
&=& 5.3\left(\frac{B_{\rm s}}{10^{12}~{\rm G}}\right)^{4/7}\left(\frac{R}{12~\rm km}\right)^{12/7}\left(\frac{M}{1.4~M_{\odot}}\right)^{-1/7} \nonumber\\
& &\times\left(\frac{\dot{M}}{10^{-6}~M_{\odot}~{\rm s^{-1}}}\right)^{-2/7}\ {\rm km},
\label{eq:rm}
\end{eqnarray}
where $B_{\rm s}$, $R$, $M$, and $\dot{M}$ are
the surface magnetic dipole field strength, radius, mass, and accretion rate of the NS, respectively,
and the corotation radius
\begin{eqnarray}
r_{\rm c} &=& \left(\frac{GM}{\Omega_{\rm s}^2}\right)^{1/3}=\left(\frac{GMP_{\rm s}^2}{4\pi^2}\right)^{1/3} \nonumber\\
&=&7.8\times10^3\left(\frac{M}{1.4~M_{\odot}}\right)^{1/3}\left(\frac{P_{\rm s}}{10~\rm s}\right)^{2/3}\ {\rm km},
\label{eq:rc}
\end{eqnarray}
where $\Omega_{\rm s}=2\pi/P_{\rm s}$ and $P_{\rm s}$
are the angular velocity and spin period of the NS' shell, respectively.
One additional key radius is the light cylinder radius,
\begin{eqnarray}
r_{\rm lc}=c/\Omega_{\rm s}=4.8\times10^5\left(\frac{P_{\rm s}}{10~\rm s}\right)\ {\rm km}.
\label{eq:r_lc}
\end{eqnarray}
One expects that in the case of $r_{\rm m}<r_{\rm c}<r_{\rm lc}$ for a normal NS in an NSWD merger
with $B_{\rm s}=10^{12}~{\rm G}$, $M=1.4~M_{\odot}$, $R=12~{\rm km}$, and $P_{\rm s}=10~{\rm s}$,
as well as an accretion rate $\dot{M}=10^{-6}~M_{\odot}~{\rm s}^{-1}$,
disk material is column-accreted onto the NS' surface and leads to the NS' shell to spin up \citep{frank92}.
Following \cite{piro11} and \cite{dai06}, the time-dependent angular velocity for the NS' shell
can be solved by
\begin{eqnarray}
	I_{\rm s}\frac{d\Omega_{\rm s}}{dt} = N_{\rm acc}-N_{\rm dip}-N_{\rm mag},
	\label{eq:Is}
\end{eqnarray}
where $I_{\rm s}$ is the moment of inertia of the shell,
and (1) $N_{\rm acc}$ is the accretion torque described by,
when $r_{\rm m}<R$ for a normal NS from Equation (\ref{eq:rm}),
\begin{equation}
N_{\rm acc}=\left(1-\frac{\Omega_{\rm s}}{\Omega_{\rm K}}\right)(GMR)^{1/2}\dot{M},
\label{eq:Nacc}
\end{equation}
where $\Omega_{\rm K}=\left(GM/R^{3}\right)^{1/2}$; (2) $N_{\rm dip}$
is the magnetic dipole radiation torque for accreting NSs with $r_{\rm m}<r_{\rm lc}$,
enhanced over the standard dipole torque by a factor of $(r_{\rm lc}/r_{\rm m})^2>1$
due to the enhanced open magnetic field lines via the compression of the magnetosphere \citep{par16,met18}, read as
\begin{eqnarray}
N_{\rm dip}&=&\frac{2B_{\rm s}^2R^6\Omega_{\rm s}^3}{3c^3}\left(\frac{r_{\rm lc}}{r_{\rm m}}\right)^2 \nonumber\\
&=&\frac{2\Omega_{\rm s}}{3c}(B_{\rm s}R^3)^{6/7}(GM)^{2/7}\dot{M}^{4/7};
\label{eq:Ndip}
\end{eqnarray}
(3) $N_{\rm mag}$ is the magnetic torque acting between the core and shell, written as
\begin{equation}
N_{\rm mag}=\frac{2}{3}R_{\rm c}^3B_{\rm r}B_{\phi},
\label{eq:Nmag}
\end{equation}
where $B_{\rm r}=B_{\rm s}/\epsilon$ (here, $\epsilon$ is defined by the
ratio of the effective surface dipole field strength to the radial field strength)
and $B_{\phi}$ are the radial magnetic field component and toroidal field component, respectively.
They can be related to each other by
\begin{equation}
\frac{d B_{\phi}}{d t}=(\Delta\Omega)B_{\rm r}\equiv(\Omega_{\rm c}-\Omega_{\rm s})B_{\rm r},
\label{eq:dB_phi}
\end{equation}
where $\Delta\Omega$ is the differential angular velocity and $\Omega_{\rm c}$
is the NS' core angular velocity.
On the other hand, the time-dependent angular velocity for the NS' core
can be solved by\footnote{Please note that the right term of Equation (\ref{eq:Ic})
has no negative sign, differing from Equation (3) in \cite{dai06}
because the magnetic field is amplified via the differential angular velocity
resulting from the angular momentum transport of the accreting material onto the NS' shell
rather than the angular momentum loss of the core for an accreting NS system we consider here.
The same reason that the right term of Equation (\ref{eq:Is})
in this paper differs from Equation (2) in \cite{dai06}.}
\begin{eqnarray}
	I_{\rm c}\frac{d\Omega_{\rm c}}{dt} = N_{\rm mag},
	\label{eq:Ic}
\end{eqnarray}
where $I_{\rm c}$ is the moment of inertia of the core.
Through Equations (\ref{eq:Is}), (\ref{eq:dB_phi}), and (\ref{eq:Ic}), we can solve
$\Omega_{\rm s}$, $\Omega_{\rm c}$, $\Delta\Omega$, and $B_{\phi}$ as illustrated in Figure \ref{fig:solution}
via numerical calculation,
combining Equations (\ref{eq:Nacc}) and (\ref{eq:Nmag}). To obtain these results, we have also employed:
(1) typical values for a normal NS in an NSWD merger: $M=1.4~M_{\odot}$,
$R=12~{\rm km}$, initial period $P_{\rm s,0}=10~{\rm s}$, $B_{\rm s}=10^{12}~{\rm G}$,
$I_{\rm s}\cong I_{\rm c}\sim10^{45}~{\rm g~cm^2}$, $\epsilon=0.3$ \citep{dai06},
as well as an accretion rate $\dot{M}=10^{-6}~M_{\odot}~{\rm s}^{-1}$, such that the term $N_{\rm dip}$ can be ignored in comparison with $N_{\rm acc}$ even if $P_{\rm s}$ possibly reaches down to
its break-up limit $P_{\rm s,min}=0.96~{\rm ms}$ \citep{lat04}; (2) initial conditions: $\Omega_{\rm s,0}=2\pi/P_{\rm s,0}$, $\Omega_{\rm c,0}=(1+A_0)\Omega_{\rm s,0}$ (the initial angular velocity of the core $\Omega_{\rm c,0}$ should be
larger than that of the shell $\Omega_{\rm s,0}$ for a normal NS), $A_0=10^{-3}$ related to a small residual differential rotation, and $B_{\phi,0}\sim10^8~{\rm G}$; (3) boundary conditions:
$\Omega_{\rm s}\leq\Omega_{\rm s, max}=6541$ due to $P_{\rm s,min}=0.96~{\rm ms}$,
$\Omega_{\rm c}<\Omega_{\rm c,max}=c/R_{\rm c}$,
and $B_{\phi}<B_{\phi,\rm max}=10^{17}~{\rm G}$ because of the buoyancy effect.
From Figure \ref{fig:solution}, we can acquire the following:
\begin{itemize}
\item The top panel shows that the angular velocity of the shell gradually increases up to its limit at about 200 s and lasts this lasts until the end of the dynamical process. While the angular velocity of the core reverses (rotating in an opposite direction) at about 50 s and its absolute value rapidly rises to a very large value $c/R_{\rm c}$\footnote{The core angular velocity $\Omega_{\rm c}$ that exceeds the break-up limit of the shell is reasonable since this break-up limit should not be that of the core angular velocity. Instead, its limit should be $c/R_{\rm c}$.} at around 1000 s. This is because the toroidal magnetic field $B_{\phi}$ reverses, as in the bottom panel. The evolution of the differential angular velocity $\Delta\Omega$ follows the angular velocity of the core.
\item In the bottom panel, the toroidal magnetic field $B_{\phi}$ rapidly declines to zero and then reverses before 0.1 s, its absolute value continues going up to the buoyancy limit $10^{17}~{\rm G}$ at about 150 s.
\end{itemize}
In short, these results manifest the toroidal magnetic field can be enhanced during the dynamical timescale
as long as the initial remnant NS in an NSWD merger has a small residual differential rotation.
Additionally, during the field amplification, the spin-down torque $N_{\rm dip}$
responds to the magnetic dipole radiation luminosity
\begin{eqnarray}
L_{\rm dip}&=&N_{\rm dip}\Omega_{\rm s} \nonumber \\
&=&\frac{2\Omega_{\rm s}^2}{3c}(B_{\rm s}R^3)^{6/7}(GM)^{2/7}\dot{M}^{4/7},
\end{eqnarray}
which follows the evolution of the angular velocity of the shell, as displayed in Figure \ref{fig:Ldip}.
This Poynting flux could generate a high energy (X-ray/$\gamma$-ray) transient lasting hundreds to
thousands of seconds via magnetic dissipation
with brightness up to $L_{\rm dip,peak}\sim10^{46}~{\rm erg~s^{-1}}$,
which is likely similar to an X-ray transient source named CDF-S XT2 discovered by \cite{xue19}.

Besides the differential rotation, \cite{dun92} and \cite{thom93} suggested that
a key parameter for the success of $\alpha$--$\omega$ dynamo is the Rossby number $R_O$
relevant to convection (the ratio of the rotation period to
the convective overturn time).
An efficient dynamo result needs $R_O\sim10(P/10~{\rm ms})(F/10^{39}~{\rm erg~cm^{-2}~s^{-1}})^{1/3}\leq O(1)$
(where $F$ is the entropy-driven convection heat flux).
This type of convection usually occurs in a nascent NS left behind the collapse of a massive star,
a BNS/BWD merger, or an AIC of a WD with a negative radial entropy gradient
from the interior to the outer layers \citep{thom93},
but should not occur in the old NS in an NSWD system.
However, for the accreting NS in an NSWD merger, its surface should be covered
by an accreting envelope with tidal WD debris via magnetically channeled accretion.
Under this condition,
the accretion flow can produce heat radiation due to
the shock heating \citep[see the appendix in][]{piro11},
its temperature is $T_{\rm sh}\sim8\times10^9~{\rm K}$, if $M=1.4~M_{\odot}$,
$R=12~{\rm km}$, $B_{\rm s}=10^{12}~{\rm G}$,
and $\dot{M}=10^{-6}~M_{\odot}~{\rm s}^{-1}$ are
considered \citep[for a detailed derivation please refer to Equation (A1) in][]{zhong19}.
Therefore, the shock heat cannot escape via neutrino cooling (since a low temperature cannot induce neutrino emission) or
photon diffusion (since photons are trapped and advected due to the high accretion rate).
Throughout the large and radiatively inefficient accreting envelope,
convection may be an important source of outward energy transport \citep[e.g.][]{quat00}.
In this case, the energy flux due to convection in the absence of bulk of motion or
angular momentum transport should be from gravitational potential energy flux
$F_{\rm c}\sim GM\dot{M}/(4\pi R^3)$.
If the magnetic field of the NS can be enhanced by this convection, the magnetic field
could reach to its buoyancy value $10^{17}~{\rm G}$
(please note that $B_{\rm c}^2/(8\pi)<F_{\rm c}t_{\rm dyn}\sim [GM\dot{M}/(4\pi R^3)]t_{\rm dyn}$
and thus $B_{\rm c}<1.0\times10^{20}~{\rm G}$)
under the parameter values $M=1.4~M_{\odot}$, $R=12~{\rm km}$, $\dot{M}\sim10^{-6}~M_{\odot}~{\rm s}^{-1}$,
and $t_{\rm dyn}=3\times10^4~{\rm s}$, although the real circumstance could be more complex.

Differing from the differential rotation and convection processes that occur in the accreting NS of
the NSWD merger, an alternative potential process, MRI,
possibly occurs in the disk due to the presence of shear,
like the disk dynamo in common envelope events for the formation of
highly magnetic WDs \citep{tout08,nor11}.
This process amplifies the disk field, and the field then would be conveyed to the surface of the NS
along with the radiatively inefficient magnetically channeled accretion flow.
The disk dynamo for WDs in the super-Eddington regime \citep{nor11} should also be suitable for
the accretion disk in the NSWD merger. Accordingly, based on
\cite{bla01} and \cite{nor11}, the mean toroidal field via the MRI in the disk at
radius $r$ is estimated by
\begin{eqnarray}
\bar{B}_{\phi}&\sim&\left(\frac{\dot{M}\Omega_{\rm K}}{r}\frac{r}{H}\right)^{1/2} \nonumber \\
&=&6\times10^{12}\left(\frac{\dot{M}}{10^{-6}~M_{\odot}~{\rm s}^{-1}}\right)^{1/2}\left(\frac{M}{1.4~M_{\odot}}\right)^{1/4} \nonumber \\
&\times&\left(\frac{r}{12~{\rm km}}\right)^{-5/4}\left(\frac{r/H}{2}\right)^{1/2}\ {\rm G},
\end{eqnarray}
where $\Omega_{\rm K}=(GM/r^3)^{1/2}$ is the Keplerian rate the disk orbits, $H$
is the isothermal scale height of the disk. This process can enhance the magnetic field
of the NS but could not result in a magnetar-like field.

\subsection{Fossil Field}
\label{subsec:fossil}
The second scenario that we consider is the magnetic flux conservation--fossil field scenario \citep{fer06,fer08},
which signifies that the magnetic fields of the NSs and/or the WDs in NSWD binaries
should be stronger than normal NSs and/or WDs. Although it is hard to imagine that the progenitors
in binary compact stars possess a very strong magnetic field since such strong magnetic fields
likely decay on much shorter timescales $\sim10^{4-5}~{\rm yr}$ \citep{hey98,har06}
than the merger lifetime. However, there may be some speculating clues.
For instance, the precursor of GRB 090510 is likely related to a magnetar-like
magnetic field \citep[$B>10^{15}~{\rm G}$ in][]{tro10} of the NS in the progenitors
if the precursor stems from the magnetospheric interaction of the NSs.
Furthermore, the merger lifetimes can also be much shorter than the inspiral times for
a sizable fraction of double NS mergers in some population synthesis models \citep{bel02,bel06}.
After the merger lifetimes, the magnetic fields should have decayed by only a factor of a few,
as illuminated in \cite{tro10}. This result should be suitable at least for
a small fraction (a few times 0.1\%) of NSWD mergers,
since the supernova producing NS precedes the NSWD merger
(in which the WD forms before the NS for the majority of NSWD mergers)
by less than 100 years, as suggested in \cite{too18}.

Whether or not these scenarios can enhance the magnetic fields of final NSs
post NSWD mergers still lacks evidence both in numerical simulations and observations.
Accordingly, for the problem of the fate of the remnants' magnetic fields post NSWD mergers,
deep and complex exploration and magnetohydrodynamics simulations
for the magnetic field evolution of the remnants are required.

\section{Explanations of Observational Properties of FRB 180924-like Bursts}
\label{sec:explanation}
If FRB 180924-like bursts can be produced from magnetars born in NSWD mergers,
this NSWD channel should be able to explain all of the observational properties of this FRB population,
as well as the event rate, host galaxy and offset, and circumburst environment.
Due to the similarities between the NSWD channel and BNS/BWD/AIC channels, we mainly follow \cite{mar19}
to analyze and discuss this NSWD channel, which is shown as follows.

(1) {\em Active Lifetime.}
\label{subsec:lifetime}
The mass of magnetars formed from NSWD mergers, $M_{\rm mag}$, could be smaller than
or close to the maximal mass of a non-rotating NS $M_{\rm TOV}$,
given a critical WD mass $M_{\rm WD,crit}=0.37~M_{\odot}$ or $M_{\rm WD,crit}=0.2~M_{\odot}$
for unstable mass transfer, and a canonical NS mass $M_{\rm NS}=1.4~M_{\odot}$.
However, these magnetars should have a lower mass than those born in the BNS channel,
but likely a higher mass than those born in SLSNe/LGRBs or BWD/AIC channels.
Moreover, the mass $M_{\rm mag}$ should also exceed or approach to the threshold mass
for the onset of direct or modified URCA neutrino cooling \citep{belo16}.
As pointed out by \cite{mar19}, such magnetars may possess sufficiently high central densities
(or high temperatures) to activate URCA cooling in their cores.
Otherwise, their magnetic dissipation in the core is caused predominately
by ambipolar diffusion \citep{gold92,thom96}, which is sensitive to the core temperature.
Since the core temperature depends on its URCA cooling at early times of magnetar formation,
their magnetic activity timescale in the direct URCA cooling (high-mass NS)
and modified URCA cooling (normal-mass NS) can be estimated as
$t_{\rm mag}\sim B_{16}^{-1}L_{5}^{3/2}20~\rm{yr}$ ($700B_{16}^{-1.2}L_{5}^{1.6}~\rm{yr}$)
for high-mass NS \citep[normal-mass NS; see Equation (33) of][]{belo16}.
This would correspond to a magnetic energy dissipation with an average luminosity
$L_{\rm mag}\sim 5\times10^{40} B_{16}^{3}~\rm{erg}~\rm{s}^{-1}$ ($10^{39}B_{16}^{3.2}~\rm{erg}~\rm{s}^{-1}$)
for high-mass NS \citep[normal-mass NS; see Equation (2) of][]{mar19},
which is just lower than the peak luminosities of FRBs by two to four orders of magnitude \citep{zhang18}.

During the active lifetimes of magnetars, they therefore produce $\sim 100$ repeating bursts
resulting from relativistic blast waves caused by giant flares with luminosity
higher than $L_{\rm mag}$ by several orders of magnitude, enough to satisfy
the event rate of FRBs \citep{ravi19b}. Because the active lifetimes are several tens
to several hundreds of years, the ``dark period'' between bursts can average years
to several tens of years; there should be have different local DMs between bursts
even though they stem from the same source. Under this condition,
we might regard them as different bursts from different sources rather than
repeating bursts from the same source, e.g., two possible cases: FRBs 110220 and 140514 \citep{piro17},
and FRBs 160920 and 170606. Therefore,
it is easy to understand that repeating bursts of FRB 180924 have not been detected
during a relatively short follow-up observation.
This is also supported by the bright FRB 171019 followed by faint bursts \citep{kumar19}.

(2) {\em Burst Transparency.}
\label{transparency}
In the framework of flaring magnetars, bursts can escape only when the surrounding material
is free-free transparent for radio frequency $\sim {\rm GHz}$.
Similar to BNS/BWD mergers and AIC, there should also be ejecta surrounding
the ``renascent'' magnetars for the NSWD channel, which may also give rise to
observable explosive transients. The ejecta consists of the WD debris disk
and the accretion-driven outflow with velocity extending up to $\sim3\times10^4~{\rm km~s^{-1}}=0.1c$,
with overall low mass of $0.01-0.1~M_{\odot}$ \citep{zen19b}.
If the free-free optical depth of ejecta for which the temperature
and ionization state are governed by photo-ionization due to spin-down of
the ``renascent'' magnetar, as handled in \cite{mar19},
the free-free transparency time could be $t_{\mathrm{ff}} \propto M_{\mathrm{ej}}^{2 / 5} v_{\mathrm{ej}}^{-1}$
for a fixed ionization fraction \citep[see Equation (18) in][]{mar18a}.
This result should be comparable to BNS mergers,
e.g., $M_{\rm ej}\sim0.05~M_{\odot}$ and $v_{\rm ej}\sim0.2c$ inferred from kilonova emission
accompanying GW170817 \citep{cow17,kase17,vill17}.
Hence, FRBs can pass through the ejecta quickly and escape in just about
a few weeks to months post magnetar formation,
compared with $t_{\rm ff}\sim10-100$ yr for SLSNe \citep{mar18a}.

(3) {\em Circumburst DM.}
\label{subsec:DM}
The circumburst DM contribution of ejecta to the burst should be akin to that of BNS/BWD/AIC channels,
can be calculated by \citep{mar19}
\begin{equation}
\mathrm{DM}_{\mathrm{ej}} \approx \frac{3 M_{\mathrm{ej}}}{8 \pi m_{\mathrm{p}}\left(v_{\mathrm{e} j} t\right)^{2}} \approx 5~\mathrm{pc}~\mathrm{cm}^{-3} M_{\mathrm{ej},-1} \beta_{\mathrm{ej}}^{-2} t_{\mathrm{yr}}^{-2},
\label{DM_ej}
\end{equation}
where $M_{\mathrm{ej},-1}=M_{\rm ej}/0.1~M_{\odot}$, $\beta_{\mathrm{ej}}=v_{\mathrm{ej}}/c$
and $t_{\rm yr}=t/1~{\rm year}$. For ejecta post NSWD mergers,
$M_{\rm ej}\sim0.01-0.1~M_{\odot}$ and $v_{\mathrm{ej}}\sim0.1c$,
so ${\rm DM_{ej}}\sim50-500~\mathrm{pc}~\mathrm{cm}^{-3}~t_{\mathrm{yr}}^{-2}$.
If the radio frequency is transparent after one month, ${\rm DM_{ej}}$ would decrease to $5\times10^{3-4}~\mathrm{pc}~\mathrm{cm}^{-3}$. For the case FRB 180924,
it has a mean contribution by host galaxy ${\rm DM_{host}}~30-81~\mathrm{pc}~\mathrm{cm}^{-3}$ \citep{ban19}.
The contribution by its ejecta should be smaller than ${\rm DM_{host}}$.
If so, FRB 180924 should escape from the ejecta at least one year after the magnetar is born in the NSWD channel,
in which time the radio frequency is already transparent.
Note that the most repeating FRBs have a nearly invariable DM
for long-term observations \citep{spi16,chime19a,chime19b,kumar19,chime20}
signifying that the DM contribution from their ejecta has declined close to zero
several years to several hundreds of years after their magnetar formation.

(4) {\em RM.}
\label{subsec:RM}
According to \cite{mar18a}, the maximal contribution to the RM is primarily caused by
a nebula in which the cooled electrons and magnetic field injected by magnetar flares
in the distant past and confined by the SN ejecta \citep{belo17,mar18b}.
It is given by
\begin{equation}
\begin{aligned} \mathrm{RM}=& \frac{e^{3}}{2 \pi m_{\mathrm{e}}^{2} c^{4}} \int n_{\mathrm{e}} B_{\|} \mathrm{d} s \approx \frac{3 e^{3}}{8 \pi^{2} m_{\mathrm{e}}^{2} c^{4}} \frac{N_{\mathrm{e}} B_{\mathrm{n}}}{R_{\mathrm{n}}^{2}}\left(\frac{\lambda}{R_{\mathrm{n}}}\right)^{1 / 2}, \end{aligned}
\end{equation}
where the total number of electrons in the nebula
\begin{equation}
N_{\mathrm{e}}=\xi E_{\rm mag},
\end{equation}
the magnetic field strength in the nebula
\begin{equation}
B_{\mathrm{n}} \approx\left(\frac{6 \epsilon_{B} E_{\mathrm{mag}}{\rm abs}(\alpha-1)}{R_{\mathrm{n}}^{3}}\right)^{1 / 2}\left(\frac{t}{t_{\mathrm{mag}}}\right)^{(1-\alpha) / 2},
\end{equation}
and the nebula size $R_{\rm n}$ is set by the outer ejecta radius
\begin{equation}
R_{\rm n}=v_{\rm ej}t,
\end{equation}
and $\lambda$ is correlation length-scale of the magnetic field in the nebula,
$\epsilon_{\rm B}$ is the ratio of the magnetic energy in the nebula to
the magnetic energy injected in relativistic particles over an expansion time $t$,
$\alpha$ is the decay index related to the average magnetic luminosity of the magnetar,
and $\xi$ is the average ratio of the number of ejected baryons
to the released magnetic energy \citep{belo17}.
If given $\epsilon_{\rm B}=0.1$, $\xi=\xi_{\rm max}\approx4\times10^3~{\rm erg}^{-1}$,
$E_{\rm mag}\approx3\times10^{49}~{\rm erg}$ \citep{belo16}, $v_{\rm ej}=0.1c$,
$\lambda\sim R_{\rm n}$, and $\alpha=0$, motivated
by magnetic-dissipation-powered FRB models \citep{mar18a} for
a magnetar formed from the NSWD channel, its RM
is just $\sim12~{\rm rad~m^{-2}}$ at the time $t\sim10^{-1}t_{\rm mag}$,
assuming $t_{\rm mag}\sim100~{\rm yr}$.
Moreover, its RM decreases with time.
These results are generally consistent with \cite{mar19}
and the RM observation of FRB 180924 \citep{ban19}.

(5) {\em Persistent Radio Source.}
\label{subsec:persisitent}
The persistent radio emission arising from the NSWD channel should also be analogous to that
from BNS/BWD/AIC channels, due to the similar properties of their nebula and ejecta.
Accordingly, there is no evidence for persistent radio emission in FRB 180924
that can be easily understood using synchrotron radiation in the nebula confined by the ejecta,
based on Figure 3 of \cite{mar19}.

(6) {\em Host Galaxy.}
\label{subsec:host}
\cite{met12} suggested that NSWD mergers involving pure-He WDs
could be related to faint type Ib Ca-rich SNe, which mostly explode in early-type galaxies
and old environments \citep{pere10,kasl12,lyman13}.
On the contrary, those mergers relevant to C/O or hybrid C/O/He WDs
are likely associated with the transients
most similar to SNe Ic \citep{too18,zen19b}. Moreover, \cite{too18} showed that
only a small fraction are expected to be found in early-type elliptical/S0 galaxies,
while a large subset of NSWD mergers are most likely to be found in late-type, disk,
and star-forming galaxies since the delay time distribution peaks at early times ($<1-2$ Gyr).
This is because they argued that hybrid WD mergers are more common than pure He WD mergers.
They also obtained that the offsets of NSWD mergers,
depending on the stellar density of host galaxies, could range from small offsets
due to a low escape velocity for those in dwarf galaxies to very large offsets of
up to a few hundred kiloparsecs for those including NS natal kicks.
However, it is still possible that some Ca-rich transients originate from He WD mergers,
while more massive NSWD mergers give rise to some kind of
fast-evolving Ic-like transients \citep{mar16,fern19}.
Therefore, we can see that the host galaxies of NSWD mergers are in a large range.
In this case, there should be a subset of galaxies hosted by NSWD mergers
to satisfy the properties of host galaxies of FRB 180924-like bursts.
Thanks to the large range in local environments and host galaxies, however,
this channel can also account for the properties of local environment and
host galaxy of FRB 180916.J0158+65, i.e., this FRB locates at a star-forming region
in a massive spiral galaxy \citep{marc20}.

(7) {\em Event Rate.}
\label{subsec:rate}
The volumetric event rate of NSWD mergers is in a range of
$(0.5-1)\times10^4~{\rm Gpc}^{-3}{\rm yr}^{-1}$ in the local universe,
which is $\sim2.5\times10^3-1\times10^4$ times
more than that of the observed LGRBs \citep{thom09} but roughly lower than
that of FRBs by one order of magnitude \citep{nich17}.
\cite{kho19} obtained a roughly consistent result that the total rate of NSWD mergers
is $\sim850~{\rm sky}^{-1}{\rm day}^{-1}$ using cosmic star formation history from \cite{mad14},
which is approximately lower than the rate of FRBs
$\sim10^{3-4}~{\rm sky}^{-1}{\rm day}^{-1}$ \citep{cor19} by one order of magnitude.
However, the fraction of NSWD mergers generating magnetars is very unclear.
If this fraction is comparable to that of BNS mergers, i.e., 3\%,
as estimated in \cite{nich17}, the rate of magnetars formed from NSWD mergers
is approximately $150-300~{\rm Gpc}^{-3}{\rm yr}^{-1}$ which
is comparable to the overall rate of millisecond magnetars born in SLSNe/LGRBs
and SGRBs---BNS channels, i.e., few $10-100~{\rm Gpc}^{-3}{\rm yr}^{-1}$ in \cite{nich17}.
If this is the case, magnetars formed from the NSWD channel can contribute to
at least a subset of FRB 180924-like bursts.
Due to a large uncertainty of the magnetar formation rate in the NSWD channel,
magnetars formed from this channel are also required to emit several bursts over their lifetimes,
especially their active lifetimes, if all of FRB 180924-like bursts result from magnetars,
based on \cite{ravi19b}.

\section{Summary and Discussion}
\label{sec:summary}
Assuming there are two FRB populations: FRB 121102-like bursts arise from magnetars
born in SLSNe/LGRBs channels while FRB 180924-like bursts arise from magnetars
born in BNS/BWD/AIC channels, we have investigated whether FRB 180924-like bursts
could also arise from magnetars formed from the NSWD channel, i.e.,
(1) whether magnetars can be formed from NSWD mergers with unstable mass transfer,
and (2) if they can indeed, whether flaring magnetars formed from this channel
can explain the observations of FRB 180924-like bursts such as their own characteristics,
local environments, host galaxies, and event rate.
We explored the first question and speculated that there are two possible scenarios
to produce strongly magnetized ``renascent'' NSs from NSWD mergers.
The first scenario is magnetic field amplification
by a vigorous $\alpha$--$\omega$ dynamo acting on the accreting NS surrounded by
a massive extended hot disk composed of WD debris during the mass transfer process.
We performed a preliminary calculation and showed that the magnetic field of the final NS
could be enhanced via the dynamo induced by differential rotation and convection
in/on the accreting NS,
as well as the MRI in the disk.
The second scenario is magnetic flux conservation of a fossil field. This scenario could contribute to a small fraction of NSWD binaries
in which the NSs are strongly magnetized and remain their magnetic fields before coalescence.
Whether or not these scenarios can give rise to magnetars post NSWD mergers
still requires evidence from both numerical simulations and observations.
As a result, the magnetic field evolution of the remnants requires some deep and complex
exploration and magnetohydrodynamics simulations.

In any case, if the NSWD channel can create magnetars,
it could produce FRB 180924-like bursts and account for their properties over an active lifetime,
burst transparency, circumburst DM, RM, persistent radio source, host galaxy,
and event rate within the framework of flaring magnetars
because the ejecta, local environments, and host galaxies of the final remnants from
this channel resemble those of BNS/BWD/AIC channels.
Otherwise, within a large range in local environment and host galaxy,
this channel can also account for the observational properties of FRB 180916.J0158+65 \citep{marc20}:
not only its properties similar to FRB 180924 such as circumburst DM, RM, and persistent radio source
because of the similar ejecta,
but also its local environment and host galaxy differing from FRB 180924.

In the future, an evident association between FRBs and magnetars formed from NSWD mergers
should need an association of transients most similar to faint type Ib Ca-rich SNe \citep{met12}
or SNe Ic \citep{zen19b}, gravitational waves from NSWD during the inspiral
and merger phase detected by eLISA or even aLIGO/Virgo \citep{pas09},
and FRBs, if such bursts are indeed produced from flaring magnetars.

\acknowledgments
We would like to thank the referee for his/her very careful and helpful
comments and suggestions that have allowed us to improve the
presentation of this manuscript significantly. This work was supported by the National Key
Research and Development Program of China (grant No.
2017YFA0402600) and the National Natural Science Foundation
of China (grant No. 11833003).


\begin{thebibliography}{}
\expandafter\ifx\csname natexlab\endcsname\relax\def\natexlab#1{#1}\fi
\providecommand{\url}[1]{\href{#1}{#1}}
\providecommand{\dodoi}[1]{doi:~\href{http://doi.org/#1}{\nolinkurl{#1}}}
\providecommand{\doeprint}[1]{\href{http://ascl.net/#1}{\nolinkurl{http://ascl.net/#1}}}
\providecommand{\doarXiv}[1]{\href{https://arxiv.org/abs/#1}{\nolinkurl{https://arxiv.org/abs/#1}}}

\bibitem[{{Balbus} \& {Hawley}(1998)}]{bal98}
{Balbus}, S.~A., {Hawley}, J.~F. 1998, Reviews of Modern Physics, 70, 1,
  \dodoi{10.1103/RevModPhys.70.1}

\bibitem[{{Bannister} {et~al.}(2019){Bannister}, {Deller}, {Phillips},
  {Macquart}, {Prochaska}, {Tejos}, {Ryder}, {Sadler}, {Shannon}, {Simha},
  {Day}, {McQuinn}, {North-Hickey}, {Bhandari}, {Arcus}, {Bennert}, {Burchett},
  {Bouwhuis}, {Dodson}, {Ekers}, {Farah}, {Flynn}, {James}, {Kerr}, {Lenc},
  {Mahony}, {O'Meara}, {Os{\l}owski}, {Qiu}, {Treu}, {U}, {Bateman}, {Bock},
  {Bolton}, {Brown}, {Bunton}, {Chippendale}, {Cooray}, {Cornwell}, {Gupta},
  {Hayman}, {Kesteven}, {Koribalski}, {MacLeod}, {McClure-Griffiths},
  {Neuhold}, {Norris}, {Pilawa}, {Qiao}, {Reynolds}, {Roxby}, {Shimwell},
  {Voronkov}, \& {Wilson}}]{ban19}
{Bannister}, K.~W., {Deller}, A.~T., {Phillips}, C., {et~al.} 2019, Science,
  365, 565, \dodoi{10.1126/science.aaw5903}

\bibitem[{{Belczynski} {et~al.}(2002){Belczynski}, {Kalogera}, \&
  {Bulik}}]{bel02}
{Belczynski}, K., {Kalogera}, V., \& {Bulik}, T. 2002, \apj, 572, 407,
  \dodoi{10.1086/340304}

\bibitem[{{Belczynski} {et~al.}(2006){Belczynski}, {Perna}, {Bulik},
  {Kalogera}, {Ivanova}, \& {Lamb}}]{bel06}
{Belczynski}, K., {Perna}, R., {Bulik}, T., {et~al.} 2006, \apj, 648, 1110,
  \dodoi{10.1086/505169}

\bibitem[{{Beloborodov}(2017)}]{belo17}
{Beloborodov}, A.~M. 2017, \apjl, 843, L26, \dodoi{10.3847/2041-8213/aa78f3}

\bibitem[{{Beloborodov}(2019)}]{belo19}
---. 2019, arXiv e-prints, arXiv:1908.07743.
\newblock \doarXiv{1908.07743}

\bibitem[{{Beloborodov} \& {Li}(2016)}]{belo16}
{Beloborodov}, A.~M., \& {Li}, X. 2016, \apj, 833, 261,
  \dodoi{10.3847/1538-4357/833/2/261}

\bibitem[{{Blackman} {et~al.}(2001)}]{bla01}
{Blackman}, E.~G., {Frank}, A., {Welch}, C. 2001, \apj, 546, 288,
  \dodoi{10.1086/318253}

\bibitem[{{Bobrick} {et~al.}(2017){Bobrick}, {Davies}, \& {Church}}]{bob17}
{Bobrick}, A., {Davies}, M.~B., \& {Church}, R.~P. 2017, \mnras, 467, 3556,
  \dodoi{10.1093/mnras/stx312}

\bibitem[{{Champion} {et~al.}(2016){Champion}, {Petroff}, {Kramer}, {Keith},
  {Bailes}, {Barr}, {Bates}, {Bhat}, {Burgay}, {Burke-Spolaor}, {Flynn},
  {Jameson}, {Johnston}, {Ng}, {Levin}, {Possenti}, {Stappers}, {van Straten},
  {Thornton}, {Tiburzi}, \& {Lyne}}]{cham16}
{Champion}, D.~J., {Petroff}, E., {Kramer}, M., {et~al.} 2016, \mnras, 460,
  L30, \dodoi{10.1093/mnrasl/slw069}

\bibitem[{{Chatterjee} {et~al.}(2017){Chatterjee}, {Law}, {Wharton},
  {Burke-Spolaor}, {Hessels}, {Bower}, {Cordes}, {Tendulkar}, {Bassa},
  {Demorest}, {Butler}, {Seymour}, {Scholz}, {Abruzzo}, {Bogdanov}, {Kaspi},
  {Keimpema}, {Lazio}, {Marcote}, {McLaughlin}, {Paragi}, {Ransom}, {Rupen},
  {Spitler}, \& {van Langevelde}}]{chat17}
{Chatterjee}, S., {Law}, C.~J., {Wharton}, R.~S., {et~al.} 2017, \nat, 541, 58,
  \dodoi{10.1038/nature20797}

\bibitem[{{CHIME/FRB Collaboration} {et~al.}(2019a){CHIME/FRB Collaboration},
  {Amiri}, {Bandura}, {Bhardwaj}, {Boubel}, {Boyce}, {Boyle}, {. Brar},
  {Burhanpurkar}, {Cassanelli}, {Chawla}, {Cliche}, {Cubranic}, {Deng},
  {Denman}, {Dobbs}, {Fandino}, {Fonseca}, {Gaensler}, {Gilbert}, {Gill},
  {Giri}, {Good}, {Halpern}, {Hanna}, {Hill}, {Hinshaw}, {H{\"o}fer},
  {Josephy}, {Kaspi}, {Landecker}, {Lang}, {Lin}, {Masui}, {Mckinven},
  {Mena-Parra}, {Merryfield}, {Michilli}, {Milutinovic}, {Moatti}, {Naidu},
  {Newburgh}, {Ng}, {Patel}, {Pen}, {Pinsonneault-Marotte}, {Pleunis},
  {Rafiei-Ravandi}, {Rahman}, {Ransom}, {Renard}, {Scholz}, {Shaw}, {Siegel},
  {Smith}, {Stairs}, {Tendulkar}, {Tretyakov}, {Vanderlinde}, \&
  {Yadav}}]{chime19a}
{CHIME/FRB Collaboration}, {Amiri}, M., {Bandura}, K., {et~al.} 2019a, \nat,
  566, 235, \dodoi{10.1038/s41586-018-0864-x}

\bibitem[{{CHIME/FRB Collaboration} {et~al.}(2019b){CHIME/FRB Collaboration},
  {Andersen}, {Bandura}, {Bhardwaj}, {Boubel}, {Boyce}, {Boyle}, {Brar},
  {Cassanelli}, {Chawla}, {Cubranic}, {Deng}, {Dobbs}, {Fandino}, {Fonseca},
  {Gaensler}, {Gilbert}, {Giri}, {Good}, {Halpern}, {Hill}, {Hinshaw},
  {H{\"o}fer}, {Josephy}, {Kaspi}, {Kothes}, {Landecker}, {Lang}, {Li}, {Lin},
  {Masui}, {Mena-Parra}, {Merryfield}, {Mckinven}, {Michilli}, {Milutinovic},
  {Naidu}, {Newburgh}, {Ng}, {Patel}, {Pen}, {Pinsonneault-Marotte}, {Pleunis},
  {Rafiei-Ravandi}, {Rahman}, {Ransom}, {Renard}, {Scholz}, {Siegel}, {Singh},
  {Smith}, {Stairs}, {Tendulkar}, {Tretyakov}, {Vanderlinde}, {Yadav}, \&
  {Zwaniga}}]{chime19b}
{CHIME/FRB Collaboration}, {Andersen}, B.~C., {Bandura}, K., {et~al.} 2019b,
  \apjl, 885, L24, \dodoi{10.3847/2041-8213/ab4a80}

\bibitem[{{CHIME/FRB Collaboration} {et~al.}(2020){CHIME/FRB Collaboration}}]{chime20}
{CHIME/FRB Collaboration}, {Fonseca}, E., {Andersen}, B.~C., {Bhardwaj}, M., et al., 2020, arXiv e-prints,
  arXiv:2001.03595.
\newblock \doarXiv{2001.03595}

\bibitem[{{Cordes} \& {Chatterjee}(2019)}]{cor19}
{Cordes}, J.~M., \& {Chatterjee}, S. 2019, \araa, 57, 417,
  \dodoi{10.1146/annurev-astro-091918-104501}

\bibitem[{{Cowperthwaite} {et~al.}(2017){Cowperthwaite}, {Berger}, {Villar},
  {Metzger}, {Nicholl}, {Chornock}, {Blanchard}, {Fong}, {Margutti},
  {Soares-Santos}, {Alexander}, {Allam}, {Annis}, {Brout}, {Brown}, {Butler},
  {Chen}, {Diehl}, {Doctor}, {Drout}, {Eftekhari}, {Farr}, {Finley}, {Foley},
  {Frieman}, {Fryer}, {Garc{\'\i}a-Bellido}, {Gill}, {Guillochon}, {Herner},
  {Holz}, {Kasen}, {Kessler}, {Marriner}, {Matheson}, {Neilsen}, {Quataert},
  {Palmese}, {Rest}, {Sako}, {Scolnic}, {Smith}, {Tucker}, {Williams},
  {Balbinot}, {Carlin}, {Cook}, {Durret}, {Li}, {Lopes}, {Louren{\c{c}}o},
  {Marshall}, {Medina}, {Muir}, {Mu{\~n}oz}, {Sauseda}, {Schlegel}, {Secco},
  {Vivas}, {Wester}, {Zenteno}, {Zhang}, {Abbott}, {Banerji}, {Bechtol},
  {Benoit-L{\'e}vy}, {Bertin}, {Buckley-Geer}, {Burke}, {Capozzi}, {Carnero
  Rosell}, {Carrasco Kind}, {Castander}, {Crocce}, {Cunha}, {D'Andrea}, {da
  Costa}, {Davis}, {DePoy}, {Desai}, {Dietrich}, {Drlica-Wagner}, {Eifler},
  {Evrard}, {Fernand ez}, {Flaugher}, {Fosalba}, {Gaztanaga}, {Gerdes},
  {Giannantonio}, {Goldstein}, {Gruen}, {Gruendl}, {Gutierrez}, {Honscheid},
  {Jain}, {James}, {Jeltema}, {Johnson}, {Johnson}, {Kent}, {Krause}, {Kron},
  {Kuehn}, {Nuropatkin}, {Lahav}, {Lima}, {Lin}, {Maia}, {March}, {Martini},
  {McMahon}, {Menanteau}, {Miller}, {Miquel}, {Mohr}, {Neilsen}, {Nichol},
  {Ogando}, {Plazas}, {Roe}, {Romer}, {Roodman}, {Rykoff}, {Sanchez},
  {Scarpine}, {Schindler}, {Schubnell}, {Sevilla-Noarbe}, {Smith}, {Smith},
  {Sobreira}, {Suchyta}, {Swanson}, {Tarle}, {Thomas}, {Thomas}, {Troxel},
  {Vikram}, {Walker}, {Wechsler}, {Weller}, {Yanny}, \& {Zuntz}}]{cow17}
{Cowperthwaite}, P.~S., {Berger}, E., {Villar}, V.~A., {et~al.} 2017, \apjl,
  848, L17, \dodoi{10.3847/2041-8213/aa8fc7}

\bibitem[{{Cumming} {et~al.}(2004){Cumming}, {Arras}, \& {Zweibel}}]{cum04}
{Cumming}, A., {Arras}, P., \& {Zweibel}, E. 2004, \apj, 609, 999,
  \dodoi{10.1086/421324}

\bibitem[{{Dai} {et~al.}(2016){Dai}, {Wang}, {Wu}, \& {Huang}}]{dai16}
{Dai}, Z.~G., {Wang}, J.~S., {Wu}, X.~F., \& {Huang}, Y.~F. 2016, \apj, 829,
  27, \dodoi{10.3847/0004-637X/829/1/27}

\bibitem[{{Dai} {et~al.}(2006){Dai}, {Wang}, {Wu}, \& {Zhang}}]{dai06}
{Dai}, Z.~G., {Wang}, X.~Y., {Wu}, X.~F., \& {Zhang}, B. 2006, Science, 311,
  1127, \dodoi{10.1126/science.1123606}

\bibitem[{{Deng} {et~al.}(2018){Deng}, {Cai}, {Wu}, \& {Liang}}]{deng18}
{Deng}, C.-M., {Cai}, Y., {Wu}, X.-F., \& {Liang}, E.-W. 2018, \prd, 98,
  123016, \dodoi{10.1103/PhysRevD.98.123016}

\bibitem[{{Duncan} \& {Thompson}(1992)}]{dun92}
{Duncan}, R.~C., \& {Thompson}, C. 1992, \apjl, 392, L9, \dodoi{10.1086/186413}

\bibitem[{{Falcke} \& {Rezzolla}(2014)}]{fal14}
{Falcke}, H., \& {Rezzolla}, L. 2014, \aap, 562, A137,
  \dodoi{10.1051/0004-6361/201321996}

\bibitem[{{Farah} {et~al.}(2018){Farah}, {Flynn}, {Bailes}, {Jameson},
  {Bannister}, {Barr}, {Bateman}, {Bhand ari}, {Caleb}, {Campbell-Wilson},
  {Chang}, {Deller}, {Green}, {Hunstead}, {Jankowski}, {Keane}, {Macquart},
  {M{\"o}ller}, {Onken}, {Os{\l}owski}, {Parthasarathy}, {Plant}, {Ravi},
  {Shannon}, {Tucker}, {Venkatraman Krishnan}, \& {Wolf}}]{far18}
{Farah}, W., {Flynn}, C., {Bailes}, M., {et~al.} 2018, \mnras, 478, 1209,
  \dodoi{10.1093/mnras/sty1122}

\bibitem[{{Fern{\'a}ndez} {et~al.}(2019)}]{fern19}
{Fern{\'a}ndez}, R., {Margalit}, B., {Metzger}, B.
2019, \mnras, 488, 259,
\dodoi{10.1093/mnras/stz1701}

\bibitem[{{Ferrario} \& {Wickramasinghe}(2006)}]{fer06}
{Ferrario}, L., \& {Wickramasinghe}, D. 2006, \mnras, 367, 1323,
  \dodoi{10.1111/j.1365-2966.2006.10058.x}

\bibitem[{{Ferrario} \& {Wickramasinghe}(2008)}]{fer08}
---. 2008, \mnras, 389, L66, \dodoi{10.1111/j.1745-3933.2008.00527.x}

\bibitem[{{Frank} {et~al.}(1992){Frank}, {King}, \& {Raine}}]{frank92}
{Frank}, J., {King}, A., \& {Raine}, D. 1992, Camb. Astrophys. Ser, {Accretion power in
  astrophysics.}, Cambridge University Press, Cambridge, England, Vol.~21

\bibitem[{{Geng} \& {Huang}(2015)}]{geng15}
{Geng}, J.~J., \& {Huang}, Y.~F. 2015, \apj, 809, 24,
  \dodoi{10.1088/0004-637X/809/1/24}

\bibitem[{{Giacomazzo} \& {Perna}(2013)}]{gia13}
{Giacomazzo}, B., \& {Perna}, R. 2013, \apjl, 771, L26,
  \dodoi{10.1088/2041-8205/771/2/L26}

\bibitem[{{Goldreich} \& {Reisenegger}(1992)}]{gold92}
{Goldreich}, P., \& {Reisenegger}, A. 1992, \apj, 395, 250,
  \dodoi{10.1086/171646}

\bibitem[{{Gu} {et~al.}(2016){Gu}, {Dong}, {Liu}, {Ma}, \& {Wang}}]{gu16}
{Gu}, W.-M., {Dong}, Y.-Z., {Liu}, T., {Ma}, R., \& {Wang}, J. 2016, \apjl,
  823, L28, \dodoi{10.3847/2041-8205/823/2/L28}

\bibitem[{{Harding} \& {Lai}(2006)}]{har06}
{Harding}, A.~K., \& {Lai}, D. 2006, Reports on Progress in Physics, 69, 2631,
  \dodoi{10.1088/0034-4885/69/9/R03}

\bibitem[{{Hessels} {et~al.}(2019){Hessels}, {Spitler}, {Seymour}, {Cordes},
  {Michilli}, {Lynch}, {Gourdji}, {Archibald}, {Bassa}, {Bower}, {Chatterjee},
  {Connor}, {Crawford}, {Deneva}, {Gajjar}, {Kaspi}, {Keimpema}, {Law},
  {Marcote}, {McLaughlin}, {Paragi}, {Petroff}, {Ransom}, {Scholz}, {Stappers},
  \& {Tendulkar}}]{hes19}
{Hessels}, J.~W.~T., {Spitler}, L.~G., {Seymour}, A.~D., {et~al.} 2019, \apjl,
  876, L23, \dodoi{10.3847/2041-8213/ab13ae}

\bibitem[{{Heyl} \& {Kulkarni}(1998)}]{hey98}
{Heyl}, J.~S., \& {Kulkarni}, S.~R. 1998, \apjl, 506, L61,
  \dodoi{10.1086/311628}

\bibitem[{{Hjellming} \& {Webbink}(1987)}]{hje87}
{Hjellming}, M.~S., \& {Webbink}, R.~F. 1987, \apj, 318, 794,
  \dodoi{10.1086/165412}

\bibitem[{{Hurley} {et~al.}(2002){Hurley}, {Tout}, \& {Pols}}]{hur02}
{Hurley}, J.~R., {Tout}, C.~A., \& {Pols}, O.~R. 2002, \mnras, 329, 897,
  \dodoi{10.1046/j.1365-8711.2002.05038.x}

\bibitem[{{Kasen} {et~al.}(2017){Kasen}, {Metzger}, {Barnes}, {Quataert}, \&
  {Ramirez-Ruiz}}]{kase17}
{Kasen}, D., {Metzger}, B., {Barnes}, J., {Quataert}, E., \& {Ramirez-Ruiz}, E.
  2017, \nat, 551, 80, \dodoi{10.1038/nature24453}

\bibitem[{{Kashiyama} {et~al.}(2013){Kashiyama}, {Ioka}, \&
  {M{\'e}sz{\'a}ros}}]{kash13}
{Kashiyama}, K., {Ioka}, K., \& {M{\'e}sz{\'a}ros}, P. 2013, \apjl, 776, L39,
  \dodoi{10.1088/2041-8205/776/2/L39}

\bibitem[{{Kasliwal} {et~al.}(2012){Kasliwal}, {Kulkarni}, {Gal-Yam}, {Nugent},
  {Sullivan}, {Bildsten}, {Yaron}, {Perets}, {Arcavi}, {Ben-Ami}, {Bhalerao},
  {Bloom}, {Cenko}, {Filippenko}, {Frail}, {Ganeshalingam}, {Horesh}, {Howell},
  {Law}, {Leonard}, {Li}, {Ofek}, {Polishook}, {Poznanski}, {Quimby},
  {Silverman}, {Sternberg}, \& {Xu}}]{kasl12}
{Kasliwal}, M.~M., {Kulkarni}, S.~R., {Gal-Yam}, A., {et~al.} 2012, \apj, 755,
  161, \dodoi{10.1088/0004-637X/755/2/161}

\bibitem[{{Katz}(2016)}]{katz16}
{Katz}, J.~I. 2016, \apj, 826, 226, \dodoi{10.3847/0004-637X/826/2/226}

\bibitem[{{Katz}(2018)}]{katz18}
---. 2018, Progress in Particle and Nuclear Physics, 103, 1,
  \dodoi{10.1016/j.ppnp.2018.07.001}

\bibitem[{{Keane} {et~al.}(2012){Keane}, {Stappers}, {Kramer}, \&
  {Lyne}}]{kean12}
{Keane}, E.~F., {Stappers}, B.~W., {Kramer}, M., \& {Lyne}, A.~G. 2012, \mnras,
  425, L71, \dodoi{10.1111/j.1745-3933.2012.01306.x}

\bibitem[{{Khokhriakova} \& {Popov}(2019)}]{kho19}
{Khokhriakova}, A.~D., \& {Popov}, S.~B. 2019, Journal of High Energy
  Astrophysics, 24, 1, \dodoi{10.1016/j.jheap.2019.09.004}

\bibitem[{{King} {et~al.}(2001){King}, {Pringle}, \& {Wickramasinghe}}]{king01}
{King}, A.~R., {Pringle}, J.~E., \& {Wickramasinghe}, D.~T. 2001, \mnras, 320,
  L45, \dodoi{10.1046/j.1365-8711.2001.04184.x}

\bibitem[{{Konar} \& {Bhattacharya}(1997)}]{konar97}
{Konar}, S., \& {Bhattacharya}, D. 1997, \mnras, 284, 311,
  \dodoi{10.1093/mnras/284.2.311}

\bibitem[{{Kumar} {et~al.}(2017){Kumar}, {Lu}, \& {Bhattacharya}}]{kumar17}
{Kumar}, P., {Lu}, W., \& {Bhattacharya}, M. 2017, \mnras, 468, 2726,
  \dodoi{10.1093/mnras/stx665}

\bibitem[{{Kumar} {et~al.}(2019){Kumar}, {Shannon}, {Os{\l}owski}, {Qiu},
  {Bhandari}, {Farah}, {Flynn}, {Kerr}, {Lorimer}, {Macquart}, {Ng},
  {Phillips}, {Price}, \& {Spiewak}}]{kumar19}
{Kumar}, P., {Shannon}, R.~M., {Os{\l}owski}, S., {et~al.} 2019, \apjl, 887,
  L30, \dodoi{10.3847/2041-8213/ab5b08}

\bibitem[{{Lattimer} \& {Prakash}(2004)}]{lat04}
{Lattimer}, J.~M., \& {Prakash}, M. 2004, Science, 304, 536,
  \dodoi{10.1126/science.1090720}

\bibitem[{{Liu}(2018)}]{liu18}
{Liu}, X. 2018, \apss, 363, 242, \dodoi{10.1007/s10509-018-3462-3}

\bibitem[{{Liu}(2020)}]{liu20}
{Liu}, X. 2020, arXiv e-prints,
  arXiv:2002.03693.
\newblock \doarXiv{2002.03693}

\bibitem[{{Lorimer} {et~al.}(2007){Lorimer}, {Bailes}, {McLaughlin},
  {Narkevic}, \& {Crawford}}]{lorimer07}
{Lorimer}, D.~R., {Bailes}, M., {McLaughlin}, M.~A., {Narkevic}, D.~J., \&
  {Crawford}, F. 2007, Science, 318, 777, \dodoi{10.1126/science.1147532}

\bibitem[{{Lu} \& {Kumar}(2018)}]{lu18}
{Lu}, W., \& {Kumar}, P. 2018, \mnras, 477, 2470, \dodoi{10.1093/mnras/sty716}

\bibitem[{{Lyman} {et~al.}(2013){Lyman}, {James}, {Perets}, {Anderson},
  {Gal-Yam}, {Mazzali}, \& {Percival}}]{lyman13}
{Lyman}, J.~D., {James}, P.~A., {Perets}, H.~B., {et~al.} 2013, \mnras, 434,
  527, \dodoi{10.1093/mnras/stt1038}

\bibitem[{{Lyubarsky}(2014)}]{lyub14}
{Lyubarsky}, Y. 2014, \mnras, 442, L9, \dodoi{10.1093/mnrasl/slu046}

\bibitem[{{Lyutikov} {et~al.}(2016){Lyutikov}, {Burzawa}, \& {Popov}}]{lyut16}
{Lyutikov}, M., {Burzawa}, L., \& {Popov}, S.~B. 2016, \mnras, 462, 941,
  \dodoi{10.1093/mnras/stw1669}

\bibitem[{{Madau} \& {Dickinson}(2014)}]{mad14}
{Madau}, P., \& {Dickinson}, M. 2014, \araa, 52, 415,
  \dodoi{10.1146/annurev-astro-081811-125615}

\bibitem[{{Marcote} {et~al.}(2020)}]{marc20}
{Marcote}, B., Nimmo K., Hessels J. W. T., {et~al.} 2020, \nat, 577, 9,
  \dodoi{10.1038/s41586-019-1866-z}

\bibitem[{{Margalit} {et~al.}(2019){Margalit}, {Berger}, \& {Metzger}}]{mar19}
{Margalit}, B., {Berger}, E., \& {Metzger}, B.~D. 2019, \apj, 886, 110,
  \dodoi{10.3847/1538-4357/ab4c31}

\bibitem[{{Margalit} \& {Metzger}(2016)}]{mar16}
{Margalit}, B., \& {Metzger}, B.~D. 2016, \mnras, 461, 1154,
  \dodoi{10.1093/mnras/stw1410}

\bibitem[{{Margalit} \& {Metzger}(2017)}]{mar17}
---. 2017, \mnras, 465, 2790, \dodoi{10.1093/mnras/stw2640}

\bibitem[{{Margalit} \& {Metzger}(2018)}]{mar18b}
---. 2018, \apjl, 868, L4, \dodoi{10.3847/2041-8213/aaedad}

\bibitem[{{Margalit} {et~al.}(2018){Margalit}, {Metzger}, {Berger}, {Nicholl},
  {Eftekhari}, \& {Margutti}}]{mar18a}
{Margalit}, B., {Metzger}, B.~D., {Berger}, E., {et~al.} 2018, \mnras, 481,
  2407, \dodoi{10.1093/mnras/sty2417}

\bibitem[{{Metzger}(2012)}]{met12}
{Metzger}, B.~D. 2012, \mnras, 419, 827,
  \dodoi{10.1111/j.1365-2966.2011.19747.x}

\bibitem[{{Metzger} {et~al.}(2017){Metzger}, {Berger}, \& {Margalit}}]{met17}
{Metzger}, B.~D., {Berger}, E., \& {Margalit}, B. 2017, \apj, 841, 14,
  \dodoi{10.3847/1538-4357/aa633d}

\bibitem[{{Metzger} {et~al.}(2018)}]{met18}
{Metzger}, B.~D., {Beniamini}, P., {Giannios}, D. 2018, \apl, 857, 95,
  \dodoi{10.3847/1538-4357/aab70c}

\bibitem[{{Metzger} {et~al.}(2019){Metzger}, {Margalit}, \& {Sironi}}]{met19}
{Metzger}, B.~D., {Margalit}, B., \& {Sironi}, L. 2019, \mnras, 485, 4091,
  \dodoi{10.1093/mnras/stz700}

\bibitem[{{Michilli} {et~al.}(2018){Michilli}, {Seymour}, {Hessels}, {Spitler},
  {Gajjar}, {Archibald}, {Bower}, {Chatterjee}, {Cordes}, {Gourdji}, {Heald},
  {Kaspi}, {Law}, {Sobey}, {Adams}, {Bassa}, {Bogdanov}, {Brinkman},
  {Demorest}, {Fernand ez}, {Hellbourg}, {Lazio}, {Lynch}, {Maddox}, {Marcote},
  {McLaughlin}, {Paragi}, {Ransom}, {Scholz}, {Siemion}, {Tendulkar}, {van
  Rooy}, {Wharton}, \& {Whitlow}}]{mich18}
{Michilli}, D., {Seymour}, A., {Hessels}, J.~W.~T., {et~al.} 2018, \nat, 553,
  182, \dodoi{10.1038/nature25149}

\bibitem[{{Murase} {et~al.}(2016){Murase}, {Kashiyama}, \& {M{\'e}sz{\'a}ros}}]{mur16}
{Murase}, K., {Kashiyama}, K., \& {M{\'e}sz{\'a}ros}, P.\ 2016, \mnras, 461, 1498,
  \dodoi{10.1093/mnras/stw1328}

\bibitem[{{Nicholl} {et~al.}(2017){Nicholl}, {Williams}, {Berger}, {Villar},
  {Alexander}, {Eftekhari}, \& {Metzger}}]{nich17}
{Nicholl}, M., {Williams}, P.~K.~G., {Berger}, E., {et~al.} 2017, \apj, 843,
  84, \dodoi{10.3847/1538-4357/aa794d}

\bibitem[{{Nomoto} \& {Kondo}(1991)}]{nomo91}
{Nomoto}, K., \& {Kondo}, Y. 1991, \apjl, 367, L19, \dodoi{10.1086/185922}

\bibitem[{{Nordhaus} {et~al.}(2011)}]{nor11}
{Nordhaus}, J., {Wellons}, S., {Spiegel}, D.~S.,
{Metzger}, B.~D., {Blackman}, E.~G. 2011, PNAS, 108, 3135,
\dodoi{10.1073/pnas.1015005108}

\bibitem[{{Parfrey} {et~al.}(2016)}]{par16}
{Parfrey}, K., {Spitkovsky}, A., {Beloborodov}, A.~M. 2016, \apj, 822, 33,
\dodoi{10.3847/0004-637X/822/1/33}

\bibitem[{{Paschalidis} {et~al.}(2011a){Paschalidis}, {Etienne}, {Liu}, \&
  {Shapiro}}]{pas11a}
{Paschalidis}, V., {Etienne}, Z., {Liu}, Y.~T., \& {Shapiro}, S.~L. 2011a,
  \prd, 83, 064002, \dodoi{10.1103/PhysRevD.83.064002}

\bibitem[{{Paschalidis} {et~al.}(2011b){Paschalidis}, {Liu}, {Etienne}, \&
  {Shapiro}}]{pas11b}
{Paschalidis}, V., {Liu}, Y.~T., {Etienne}, Z., \& {Shapiro}, S.~L. 2011b,
  \prd, 84, 104032, \dodoi{10.1103/PhysRevD.84.104032}

\bibitem[{{Paschalidis} {et~al.}(2009){Paschalidis}, {MacLeod}, {Baumgarte}, \&
  {Shapiro}}]{pas09}
{Paschalidis}, V., {MacLeod}, M., {Baumgarte}, T.~W., \& {Shapiro}, S.~L. 2009,
  \prd, 80, 024006, \dodoi{10.1103/PhysRevD.80.024006}

\bibitem[{{Perets} {et~al.}(2010){Perets}, {Gal-Yam}, {Mazzali}, {Arnett},
  {Kagan}, {Filippenko}, {Li}, {Arcavi}, {Cenko}, {Fox}, {Leonard}, {Moon},
  {Sand}, {Soderberg}, {Anderson}, {James}, {Foley}, {Ganeshalingam}, {Ofek},
  {Bildsten}, {Nelemans}, {Shen}, {Weinberg}, {Metzger}, {Piro}, {Quataert},
  {Kiewe}, \& {Poznanski}}]{pere10}
{Perets}, H.~B., {Gal-Yam}, A., {Mazzali}, P.~A., {et~al.} 2010, \nat, 465,
  322, \dodoi{10.1038/nature09056}

\bibitem[{{Petroff} {et~al.}(2019){Petroff}, {Hessels}, \& {Lorimer}}]{petr19}
{Petroff}, E., {Hessels}, J.~W.~T., \& {Lorimer}, D.~R. 2019, \aapr, 27, 4,
  \dodoi{10.1007/s00159-019-0116-6}

\bibitem[{{Petroff} {et~al.}(2016){Petroff}, {Barr}, {Jameson}, {Keane},
  {Bailes}, {Kramer}, {Morello}, {Tabbara}, \& {van Straten}}]{petr16}
{Petroff}, E., {Barr}, E.~D., {Jameson}, A., {et~al.} 2016, \pasa, 33, e045,
  \dodoi{10.1017/pasa.2016.35}

\bibitem[{{Piro} \& {Burke-Spolaor}(2017)}]{piro17}
{Piro}, A.~L., \& {Burke-Spolaor}, S. 2017, \apjl, 841, L30,
  \dodoi{10.3847/2041-8213/aa740d}

\bibitem[{{Piro} \& {Ott}(2011)}]{piro11}
{Piro}, A.~L., \& {Ott}, C.~D. 2011, \apj, 736, 108,
  \dodoi{10.1088/0004-637X/736/2/108}

\bibitem[{{Platts} {et~al.}(2019){Platts}, {Weltman}, {Walters}, {Tendulkar},
  {Gordin}, \& {Kandhai}}]{plat19}
{Platts}, E., {Weltman}, A., {Walters}, A., {et~al.} 2019, \physrep, 821, 1,
  \dodoi{10.1016/j.physrep.2019.06.003}

\bibitem[{{Popov} {et~al.}(2018){Popov}, {Postnov}, \& {Pshirkov}}]{popo18}
{Popov}, S.~B., {Postnov}, K.~A., \& {Pshirkov}, M.~S. 2018, Physics Uspekhi,
  61, 965, \dodoi{10.3367/UFNe.2018.03.038313}

\bibitem[{{Price} \& {Rosswog}(2006)}]{price06}
{Price}, D.~J., \& {Rosswog}, S. 2006, Science, 312, 719,
  \dodoi{10.1126/science.1125201}

\bibitem[{{Prochaska} {et~al.}(2019){Prochaska}, {Macquart}, {McQuinn},
  {Simha}, {Shannon}, {Day}, {Marnoch}, {Ryder}, {Deller}, {Bannister},
  {Bhandari}, {Bordoloi}, {Bunton}, {Cho}, {Flynn}, {Mahony}, {Phillips},
  {Qiu}, \& {Tejos}}]{pro19}
{Prochaska}, J.~X., {Macquart}, J.-P., {McQuinn}, M., {et~al.} 2019, Science,
  366, 231, \dodoi{10.1126/science.aay0073}

\bibitem[{{Quataert} \& {Gruzinov}(2000)}]{quat00}
{Quataert}, Eliot \& {Gruzinov}, Andrei. 2000, \apj, 539, 809,
  \dodoi{10.1086/309267}

\bibitem[{{Ravi}(2019)}]{ravi19b}
{Ravi}, V. 2019, Nature Astronomy, 3, 928, \dodoi{10.1038/s41550-019-0831-y}

\bibitem[{{Ravi} {et~al.}(2019){Ravi}, {Catha}, {D'Addario}, {Djorgovski},
  {Hallinan}, {Hobbs}, {Kocz}, {Kulkarni}, {Shi}, {Vedantham}, {Weinreb}, \&
  {Woody}}]{ravi19a}
{Ravi}, V., {Catha}, M., {D'Addario}, L., {et~al.} 2019, \nat, 572, 352,
  \dodoi{10.1038/s41586-019-1389-7}

\bibitem[{{Rosswog} {et~al.}(2003){Rosswog}, {Ramirez-Ruiz}, \&
  {Davies}}]{ross03}
{Rosswog}, S., {Ramirez-Ruiz}, E., \& {Davies}, M.~B. 2003, \mnras, 345, 1077,
  \dodoi{10.1046/j.1365-2966.2003.07032.x}

\bibitem[{{Schwab} {et~al.}(2015){Schwab}, {Quataert}, \& {Bildsten}}]{sch15}
{Schwab}, J., {Quataert}, E., \& {Bildsten}, L. 2015, \mnras, 453, 1910,
  \dodoi{10.1093/mnras/stv1804}

\bibitem[{{Schwab} {et~al.}(2016){Schwab}, {Quataert}, \& {Kasen}}]{sch16}
{Schwab}, J., {Quataert}, E., \& {Kasen}, D. 2016, \mnras, 463, 3461,
  \dodoi{10.1093/mnras/stw2249}

\bibitem[{{Spitler} {et~al.}(2016){Spitler}, {Scholz}, {Hessels}, {Bogdanov},
  {Brazier}, {Camilo}, {Chatterjee}, {Cordes}, {Crawford}, {Deneva}, {Ferdman},
  {Freire}, {Kaspi}, {Lazarus}, {Lynch}, {Madsen}, {McLaughlin}, {Patel},
  {Ransom}, {Seymour}, {Stairs}, {Stappers}, {van Leeuwen}, \& {Zhu}}]{spi16}
{Spitler}, L.~G., {Scholz}, P., {Hessels}, J.~W.~T., {et~al.} 2016, \nat, 531,
  202, \dodoi{10.1038/nature17168}

\bibitem[{{Sun} {et~al.}(2019){Sun}, {Li}, {Liu}, {L{\"u}}, {Wang}, \&
  {Zhu}}]{sun19}
{Sun}, S., {Li}, L., {Liu}, H., {et~al.} 2019, \pasa, 36, e005,
  \dodoi{10.1017/pasa.2018.51}

\bibitem[{{Tauris} {et~al.}(2013){Tauris}, {Sanyal}, {Yoon}, \&
  {Langer}}]{tau13}
{Tauris}, T.~M., {Sanyal}, D., {Yoon}, S.~C., \& {Langer}, N. 2013, \aap, 558,
  A39, \dodoi{10.1051/0004-6361/201321662}

\bibitem[{{Tendulkar} {et~al.}(2017){Tendulkar}, {Bassa}, {Cordes}, {Bower},
  {Law}, {Chatterjee}, {Adams}, {Bogdanov}, {Burke-Spolaor}, {Butler},
  {Demorest}, {Hessels}, {Kaspi}, {Lazio}, {Maddox}, {Marcote}, {McLaughlin},
  {Paragi}, {Ransom}, {Scholz}, {Seymour}, {Spitler}, {van Langevelde}, \&
  {Wharton}}]{ten17}
{Tendulkar}, S.~P., {Bassa}, C.~G., {Cordes}, J.~M., {et~al.} 2017, \apjl, 834,
  L7, \dodoi{10.3847/2041-8213/834/2/L7}

\bibitem[{{Thompson} \& {Duncan}(1993)}]{thom93}
{Thompson}, C., \& {Duncan}, R.~C. 1993, \apj, 408, 194, \dodoi{10.1086/172580}

\bibitem[{{Thompson} \& {Duncan}(1996)}]{thom96}
---. 1996, \apj, 473, 322, \dodoi{10.1086/178147}

\bibitem[{{Thompson} {et~al.}(2009){Thompson}, {Kistler}, \& {Stanek}}]{thom09}
{Thompson}, T.~A., {Kistler}, M.~D., \& {Stanek}, K.~Z. 2009, arXiv e-prints,
  arXiv:0912.0009.
\newblock \doarXiv{0912.0009}

\bibitem[{{Thorne} \& {Zytkow}(1977)}]{thor77}
{Thorne}, K.~S., \& {Zytkow}, A.~N. 1977, \apj, 212, 832,
  \dodoi{10.1086/155109}

\bibitem[{{Thornton} {et~al.}(2013){Thornton}, {Stappers}, {Bailes},
  {Barsdell}, {Bates}, {Bhat}, {Burgay}, {Burke-Spolaor}, {Champion}, {Coster},
  {D'Amico}, {Jameson}, {Johnston}, {Keith}, {Kramer}, {Levin}, {Milia}, {Ng},
  {Possenti}, \& {van Straten}}]{tho13}
{Thornton}, D., {Stappers}, B., {Bailes}, M., {et~al.} 2013, Science, 341, 53,
  \dodoi{10.1126/science.1236789}

\bibitem[{{Toonen} {et~al.}(2018){Toonen}, {Perets}, {Igoshev}, {Michaely}, \&
  {Zenati}}]{too18}
{Toonen}, S., {Perets}, H.~B., {Igoshev}, A.~P., {Michaely}, E., \& {Zenati},
  Y. 2018, \aap, 619, A53, \dodoi{10.1051/0004-6361/201833164}

\bibitem[{{Tout} {et~al.}(2008)}]{tout08}
{Tout}, C.~A., {Wickramasinghe}, D.~T., {Liebert}, J.,
{Ferrario}, L., {Pringle}, J.~E. 2008, \mnras, 387, 897,
\dodoi{10.1111/j.1365-2966.2008.13291.x}

\bibitem[{{Troja} {et~al.}(2010){Troja}, {Rosswog}, \& {Gehrels}}]{tro10}
{Troja}, E., {Rosswog}, S., \& {Gehrels}, N. 2010, \apj, 723, 1711,
  \dodoi{10.1088/0004-637X/723/2/1711}

\bibitem[{{Turolla} {et~al.}(2015){Turolla}, {Zane}, \& {Watts}}]{tur15}
{Turolla}, R., {Zane}, S., \& {Watts}, A.~L. 2015, Reports on Progress in
  Physics, 78, 116901, \dodoi{10.1088/0034-4885/78/11/116901}

\bibitem[{{Urpin} \& {Konenkov}(1997)}]{urp97}
{Urpin}, V., \& {Konenkov}, D. 1997, \mnras, 284, 741,
  \dodoi{10.1093/mnras/284.3.741}

\bibitem[{{van Haaften} {et~al.}(2012){van Haaften}, {Nelemans}, {Voss},
  {Wood}, \& {Kuijpers}}]{van12}
{van Haaften}, L.~M., {Nelemans}, G., {Voss}, R., {Wood}, M.~A., \& {Kuijpers},
  J. 2012, \aap, 537, A104, \dodoi{10.1051/0004-6361/201117880}

\bibitem[{{Villar} {et~al.}(2017){Villar}, {Guillochon}, {Berger}, {Metzger},
  {Cowperthwaite}, {Nicholl}, {Alexand er}, {Blanchard}, {Chornock},
  {Eftekhari}, {Fong}, {Margutti}, \& {Williams}}]{vill17}
{Villar}, V.~A., {Guillochon}, J., {Berger}, E., {et~al.} 2017, \apjl, 851,
  L21, \dodoi{10.3847/2041-8213/aa9c84}

\bibitem[{{Wang} {et~al.}(2016){Wang}, {Yang}, {Wu}, {Dai}, \& {Wang}}]{wang16}
{Wang}, J.-S., {Yang}, Y.-P., {Wu}, X.-F., {Dai}, Z.-G., \& {Wang}, F.-Y. 2016,
  \apjl, 822, L7, \dodoi{10.3847/2041-8205/822/1/L7}

\bibitem[{{Xue} {et~al.}(2019)}]{xue19}
{Xue}, Y.~Q., {Zheng}, X.~C., {Li}, Y., {et~al.} 2019, \nat, 568, 198,
\dodoi{10.1038/s41586-019-1079-5}

\bibitem[{{Yoon} {et~al.}(2007){Yoon}, {Podsiadlowski}, \& {Rosswog}}]{yoon07}
{Yoon}, S.~C., {Podsiadlowski}, P., \& {Rosswog}, S. 2007, \mnras, 380, 933,
  \dodoi{10.1111/j.1365-2966.2007.12161.x}

\bibitem[{{Zenati} {et~al.}(2019a){Zenati}, {Bobrick}, \& {Perets}}]{zen19a}
{Zenati}, Y., {Bobrick}, A., \& {Perets}, H.~B. 2019a, arXiv e-prints,
  arXiv:1908.10866.
\newblock \doarXiv{1908.10866}

\bibitem[{{Zenati} {et~al.}(2019b){Zenati}, {Perets}, \& {Toonen}}]{zen19b}
{Zenati}, Y., {Perets}, H.~B., \& {Toonen}, S. 2019b, \mnras, 486, 1805,
  \dodoi{10.1093/mnras/stz316}

\bibitem[{{Zhang}(2014)}]{zhang14}
{Zhang}, B. 2014, \apjl, 780, L21, \dodoi{10.1088/2041-8205/780/2/L21}

\bibitem[{{Zhang}(2016)}]{zhang16}
---. 2016, \apjl, 827, L31, \dodoi{10.3847/2041-8205/827/2/L31}

\bibitem[{{Zhang}(2017)}]{zhang17}
---. 2017, \apjl, 836, L32, \dodoi{10.3847/2041-8213/aa5ded}

\bibitem[{{Zhang}(2018)}]{zhang18}
---. 2018, \apjl, 867, L21, \dodoi{10.3847/2041-8213/aae8e3}

\bibitem[{{Zhong} {et~al.}(2019){Zhong}, {Dai}, \& {Li}}]{zhong19}
{Zhong}, S.-Q., {Dai}, Z.-G., \& {Li}, X.-D. 2019, \prd, 100, 123014,
  \dodoi{10.1103/PhysRevD.100.123014}

\end{thebibliography}


\end{document}